\documentclass[aps,pre,onecolumn,showpacs,superscriptaddress,noeprint]{revtex4-1}
\usepackage[utf8]{inputenc}
\usepackage{amsmath, amsfonts, amsthm, bbm}
\usepackage{amsthm}
\usepackage{amssymb}
\usepackage{pifont}
\usepackage{graphicx} 
\usepackage{bmpsize}
\usepackage{tikz}
\usepackage{float}

\begin{document}

\title{The Rise of Rationality in Blockchain Dynamics}

\author{Gabriele Di Antonio}
\thanks{These authors contributed equally to this work.}
 \affiliation{Centro Ricerche Enrico Fermi, Rome, Italy} 
  \affiliation{Istituto Superiore di Sanità, Rome, Italy}
 \affiliation{Università degli Studi Roma Tre, Rome, Italy}
 
\author{Gianni Valerio Vinci}
\thanks{These authors contributed equally to this work.}
 \affiliation{Istituto Superiore di Sanità, Rome, Italy}
  \affiliation{Università Roma Tor Vergata, Rome, Italy}
 
\author{Luciano Pietronero}
 \affiliation{Centro Ricerche Enrico Fermi, Rome, Italy}

\author{Marco Alberto Javarone}
\thanks{Corresponding author: marcojavarone@gmail.com}
 \affiliation{Università di Bari, Bari, Italy}
 \affiliation{Centro Ricerche Enrico Fermi, Rome, Italy}
  \affiliation{University College London - Centre for Blockchain Technologies, London, UK}

\date{\today}
\begin{abstract}
Taking informed decisions, namely acting rationally, is an individual attitude of paramount relevance in nature and human societies. 
In this work, we study how rationality spreads in a community.
To this end, through an agent-based model, we analyse the dynamics of a population whose individuals, endowed with a rational attitude controlled by a numerical parameter, play a simple game. The latter consists of multiple strategies, each associated with a given reward. 
The proposed model is then used as a benchmark for studying the behaviour of Bitcoin users, inferred by analysing transactions recorded in the Blockchain.
Remarkably, a population undergoing a sharp transition from irrational to rational attitudes shows a behavioural pattern similar to that of Bitcoin users, whose rationality showed up as soon as their cryptocurrency became worth just a few cents (USD).
To conclude, a behavioural analysis that relies on an entropy measure combined with a simple agent-based model allows us to detect the rise of rationality across a community.
Although further investigations are essential to corroborate our results, we deem the proposed approach could also get used for studying other social phenomena and behaviours.
\end{abstract}
\maketitle
\section{Introduction}\label{sec:introduction}
Human behaviour underlies the dynamics of several social and economic systems, such as crowds, financial markets, business organisations and companies, online communities, and many others~\cite{bellomo01,preis01,preis02,arnold01,alani01}. 
Often, these systems show complex interaction patterns among individuals whose analysis can unveil relevant information for explaining, and sometimes predicting, their mechanisms and behaviours~\cite{kok01,sornette01,vemula01}.
To this end, various modelling approaches, such as machine learning~\cite{kothari01}, network theory~\cite{steier01}, and agent-based models~\cite{garcia01} can be implemented, typically considering the nature of the available data and the specific goals.
The above-mentioned human behaviour, giving rise to attitudes and traits such as (anti-)conformity~\cite{galam01,perc04,javarone08}, stubbornness~\cite{galam02}, and polarisation~\cite{galam03}, can be mapped to some parameter added in these models. 
For instance, agents can be endowed with features for studying whether and how population dynamics get affected by them ---see~\cite{weron01,wen01}.
In this work, we focus on rationality, namely an attitude of particular interest to different scientific communities spanning from psychologists to philosophers~\cite{scott01,moshman01}, neuroscientists~\cite{demartino01,friston01}, economists~\cite{kahneman01,farmer01,lo01}, and computer scientists~\cite{tenenbaum01,parsons01}, to cite a few. 
Identifying and describing mental processes underlying rationality can shed light on human, and in general animal, brains and behaviours~\cite{kanai01}. Also, understanding what rationality is and its implications at a level of a community, or society, can support the development of intelligent agents (e.g. in Artificial Intelligence) and may help to assess potential threats and ethical risks. Moreover, rationality underlies equilibria observed in social dilemmas~\cite{traulsen01,perc03,moreno01,javarone04,javarone05} and offers advantages in skill-based competitions (e.g.~\cite{javarone06}).
Yet, detecting this relevant attitude in individuals or communities is all but trivial. To address this challenge, we develop an agent-based model and analyse its outcomes by an entropy measure. In doing so, the synthetic data obtained through numerical simulations can constitute a benchmark for studying real-world systems.
More in detail, the proposed agent-based model relates a population whose individuals play a simple game and, accordingly, take decisions by selecting a strategy. The rationality of these agents is introduced and controlled via a numerical parameter. The latter allows us to simulate decision processes varying the degree of rationality in the population. Notably, high rationality entails agents acting to maximise an individual gain, while low rationality adds randomness to the agents' decision process as they pay less attention to the potential return.
Then, among the real-world systems, we focus on the community of Bitcoin users, whose actions, namely Bitcoin transactions, get recorded in the Blockchain~\cite{satoshi01}.
To connect the dynamics of the proposed model with the dynamics of Bitcoin users, we map users' actions to game strategies. Therefore, in the synthetic and the real-world population, we can observe individuals taking decisions to play a game. 
Here, our goal is to compare the behaviours of agents with that of Bitcoin users to detect transitions between rationality and irrationality.
Remarkably, the outcomes of this study show a sharp transition to rational behaviours as soon as Bitcoin gets a market value of only a few cents (USD).
The remainder of the manuscript is organized as follows: Section~\ref{sec:blockchain_intro} offers a brief overview of the Blockchain and the world of cryptocurrencies. Section~\ref{sec:model} introduces the agent-based model. Then, Section~\ref{sec:results} shows results leading us to perform a comparative analysis between the Bitcoin users and the synthetic population.
Eventually, Section~\ref{sec:conclusion} discusses the main findings of this work and future developments.
\section{Blockchain and Cryptocurrencies}\label{sec:blockchain_intro}
The Blockchain~\cite{satoshi01,antonopoulos01} technology and cryptocurrencies, such as Bitcoin (BTC), Ethereum (ETH), and many others~\cite{demarzo01}, are pervading various aspects of our society. Cryptocurrencies are the main application of the Blockchain~\cite{tasca01}. That latter is a distributed ledger designed to record information according to specific rules and develops into a peculiar data structure which motivates its name.
Namely, data are collected into blocks that connect along a one-dimensional chain.
For the sake of clarity, here we use the following convention:
\begin{itemize}
\item Bitcoin (Uppercase) refers to the BTC cryptocurrency;
\item Blockchain (Uppercase) refers to the blockchain underpinning BTC;
\item blockchain (lowercase) refers to blockchains underpinning other cryptocurrencies.
\end{itemize}
Beyond any living debate, a generic cryptocurrency is a form of digital money. The main peculiarity of this asset is given by the possibility to manage transactions without the support of trusted third parties (e.g. a Bank). A blockchain, exploiting cryptographic protocols, constitutes such an innovative solution, offering protection against threats and frauds like the double-spending attack~\cite{vranken01}.
Yet, the famous Nakamoto consensus protocol, at the core of many blockchains as the one supporting BTC relies on the so-called Proof-of-Work (PoW). The PoW is a cryptographic proof required to validate cryptocurrency transactions, highly criticised for its energetic cost. 
Shortly, PoW-based blockchains need miners, i.e. particular users that collect cryptocurrency transactions (not yet recorded in these blockchains) forming the data structures we mentioned above, i.e. blocks. 
More specifically, miners aim to attach their newly formed block to the head of a blockchain to get a gorgeous reward called 'mining reward'. The latter constitutes an economic incentive to sustain a blockchain and attracts new miners. 
The mining process relates to finding a suitable random number to complete the generation of a new block. Thus, miners participate in a sort of lottery defined as the 'mining competition' granting a reward only to the winner. As soon as a mining competition ends and the new block is attached to the chain, a new competition starts.
Any attempt to include non-valid transactions into a block, by mistake or by trying a fraud, fails.
In particular, a newly formed block containing even a single non-valid transaction is not accepted by the Bitcoin network and gets cancelled out. Despite that, various attacks~\cite{conti01} are still possible. However, the strength and safety of a blockchain also rely on the computational effort required to perform these attacks.
In summary, PoW-based blockchains are safe, yet they rise concerns related to energy consumption (e.g.~\cite{keller01,gola01}) and sustainability~\cite{javarone03}.
Notwithstanding, today, Bitcoin is considered a full-fledged financial asset. 
Before entering into details of its market dynamics, let us specify the mechanism underlying transactions.
Let us assume Alice wants to send $5$ BTC to Bob. The defined amount must be available in Alice's wallet, resulting from previous Bitcoin transactions. 
Each incoming transaction constitutes a sort of coin in the receiver's wallet. Also, a transaction can be sent towards more than one wallet. 
For instance, Alice can create a $5$ BTC transaction with two outputs: one addressed to Bob, say equal to $4$ BTC, and the other equal to $1$ BTC (i.e. $5$ minus $4$), addressed to John (or to her wallet as a sort of change). 
Similarly, Alice could give $4$ US dollars to Bob by using a $5$ dollars banknote and asking for $1$ dollars back from Bob. Notice that, with cryptocurrencies, the whole transaction is designed by the sender, while the receiver constitutes a passive actor in this process.
We conclude this section by explaining why Bitcoin is especially interesting for our investigation. In the early period, i.e. right after Bitcoin's invention, it was worthless. Also, the Bitcoin community was small. For this reason, we assume that, in the beginning, Bitcoin was used quite freely and without pressure, as that one might have when dealing with a financial asset. Yet, later, as soon as BTC grew in value, we hypothesise users may have changed behaviour, becoming more careful. 
Generally, improving attention in managing a resource requires rationality. For this reason, we deem some hallmarks of rationality may hide in Bitcoin price dynamics, notably in its early period.\\
As an additional note, the first real-world transaction involving Bitcoin occurred on the 22nd of May 2010, known as Bitcoin Pizza Day~\cite{pizza_day}. That day, Laszlo Hanyecz paid $10,000$ BTC to purchase two Papa John's pizzas (yet, officially, the first transaction through the blockchain network occurred on the 12th of January 2009, but it was not about acquiring physical goods). Within a couple of months, in July 2010, the estimated price of Bitcoin grew from $0.008$ USD to $0.08$ USD, and the first large-scale Bitcoin exchange, Mt. Gox, made its appearance.
\section{Model and Data}\label{sec:model} 
The proposed agent-based model considers a population of $N$ agents whose dynamics rely on a $M$-strategy game. The latter develops through a series of iterations, requiring agents to select a strategy and receive, in turn, the corresponding payoff (or gain). 
The strategy selection process, realised by a Boltzmann-like distribution function, includes a temperature and a game payoff. Accordingly, the probability of selecting the $x$-th strategy, i.e. $\sigma_x$, reads
\begin{equation}\label{eq:strategy_selection}
    P(\sigma_x) = \frac{e^{-\beta g_x}}{\sum_i e^{-\beta g_i}}
\end{equation}
\noindent with $g_x = \frac{1}{\pi_x}$ inverse of the payoff associated with the $x$-th strategy and $\beta$ the inverse of the system temperature $T$.\\
The payoff value is drawn from an exponential curve with exponent $\lambda$, whose value is set to $1$ if not stated otherwise, which regulates the quality rank of strategies. 
Let us remark that equation~\ref{eq:strategy_selection} entails profitable strategies can be selected with a higher probability. Also, the temperature (or noise) constitutes the degree of freedom of the model we control for tuning the degree of rationality in the population. 
Thus, agents are more sensitive to the strategy payoff at low temperatures while, at high temperatures, the payoff becomes negligible in the strategy selection. 
In doing so, rational behaviours emerge as the system temperature reduces to small values, typically smaller than $1$. 
In summary, the described population has a strategy profile which evolves depending on the game parameters and can show a rich landscape of stable and meta-stable states.
For simplicity, we assume no interactions among agents. Therefore, rationality and payoff structure represent the two prominent elements of the game.\\
Finally, we describe the structure of the Blockchain dataset used to perform the comparative analysis with the outcomes of the agent-based model here presented.
\subsection*{Blockchain Data}
We consider Blockchain data generated from 2009 to 2022, available online at~\cite{blockchain_explorer}, composed of more than $700000$ blocks. A series of parameters describe a block~\cite{antonopoulos01}, yet we only consider three of them: the number of transactions, the number of inputs, and the number of outputs.
More in detail, the number of transactions per block corresponds to the total number of transactions collected in each block of the Blockchain. The number of inputs and the number of outputs refer to single transactions and allow collecting a given amount of Bitcoin and splitting them among more wallets. 
The number of inputs can be seen as a sort of collection of banknotes we put together to obtain a given amount of money. Similarly, the number of outputs can be seen as the way we manage such an amount of money, for instance, in a payment. 
For example, two banknotes of $10$ euro can be used to pay $18$ euro at a shop, receiving $2$ euro back. This simple transaction has two inputs, i.e. the two $10$ euro banknotes, and two outputs, i.e. $18$ euro and the $2$ euro change. Notice that the Bitcoin sender controls all the aspects of a transaction, including the potential change.
The described parameters have a numerical value, so blocks can easily be mapped into vectors~\cite{dai01}. Hence, we can map the whole Blockchain into a series of $3$-dimensional normalised vectors \textbf{B}. %
In light of that, the strategies of the proposed game correspond to our vectors, so each block composed of some vectors has a state defined by the total payoff. For instance, ten agents select some strategies at a given time, and these compose a block whose value/state equals the summation of the related payoffs.
It is worth highlighting that, to reduce the level of noise of the Blockchain data, we use a daily time scale by averaging over all blocks mined every $24$ hours.
Eventually, data related to the market value of Bitcoin, available online at~\cite{btc_trend}, are used to observe whether the Bitcoin users' behaviour changes in correspondence with relevant variations on the BTC value in US dollars. Thus, the signal of interest for this analysis is the BTC\slash USD.
\begin{figure}
    \centering
    \includegraphics[width=154mm]{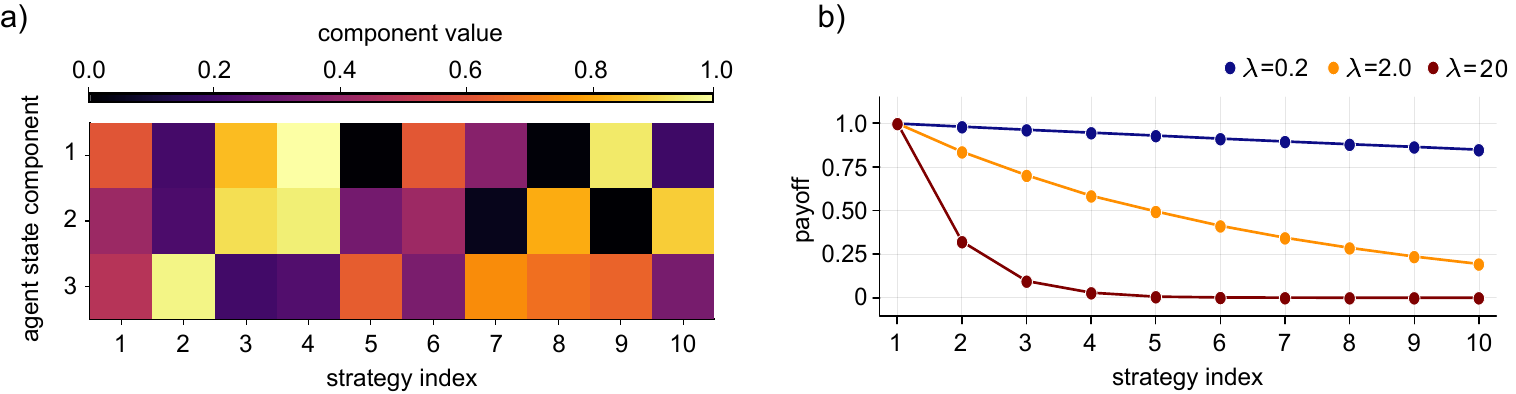}
    \caption{(a) On the top, $10$ strategies, each composed of $3$ components taking a value in the range $[0,1]$. (b) Example of payoff distribution for the $10$ generated strategies, according to an exponential distribution obtained by different values of $\lambda$. Small values of $\lambda$ lead to flattened distributions which, in turn, result in payoff homogeneity.}
    \label{fig:payoff}
\end{figure}
We remark that, for the aim of this investigation, we are not interested in predicting market trends (e.g. as in~\cite{kristoufek01,blau01,amjad01,ortu01,marchesi02,baronchelli02}) or market dynamics~\cite{kondor01,baronchelli01,tessone01,javarone01}. Notably, our goal lies in detecting a specific human attitude, i.e. rationality, and trying to capture its possible fluctuations in a community.
For the sake of clarity, we highlight that our data relates to blocks, not the activity of single individuals. Therefore, for studying the state of a block at time-step $t$, we use the following entropy measure (henceforth, referred to as local entropy)
\begin{equation}\label{eq:entropy_measure}
    S_{t;r} = - p_{t;r} \log \left( p_{t;r} \right)
\end{equation}
\noindent where $p_{t;r}$ is the probability to observe, in a time window, block states whose similarity with the block state $\textbf{B}_t$ (i.e. the block generated at the time $t$) is greater than a threshold $r$.
More in detail, the probability $p_{t;r}$ reads
\begin{equation}\label{eq:prob_entropy}
     p_{t;r} = \frac{\sum_j^K \theta\left[S_C(\textbf{B}_t,\textbf{B}_j) - r \right] }{K}
\end{equation}
\noindent where $K$ denotes the number of samples, $\theta$ represents the Heaviside step function, and $S_C$ is the cosine similarity (whose value ranges in $[-1,1]$). In this investigation, we set the similarity threshold to $r = 0.95$. 
\\
The local entropy results in high values for medium-frequency blocks, while it has small values for high-frequency blocks (i.e. common events) and low-frequency (i.e. rare events) ones. 
Assuming that a set of random agents lead to a medium-frequency block, e.g. due to brief fluctuations occurring during the observation period, irrationality is detected by a high (and fluctuating) local entropy.
\section{Results} \label{sec:results} 
The proposed model has many parameters that can affect the strategy phase of the agent population. Here, we consider only those we find more relevant and study their effect on the local entropy $S$.
We remark our population grows over time, i.e. new agents add to the system. Hence, we describe the population size by the time-dependent variable $N(t)$. For the sake of simplicity, we only consider a linear growth of agents, setting the initial value to $N(0) = 10$. 
Moreover, without loss of generality and if not otherwise stated, we consider a game with $M = 50$ strategies, labelled by numbers, e.g. $1$, $2$, $3$, and so on.
Considering the above-defined setting, we analyse some outcomes of the $M$-strategy game. This initial part of the work aims to assess whether and how this game gets affected by the system temperature. Then, we move our attention to Blockchain and Bitcoin market data, i.e. to the $BTC\slash USD$ signal in the period 2009 - 2022.
\subsection*{$M$-strategy Game}
\begin{figure}
    \centering
    \includegraphics[width=160mm]{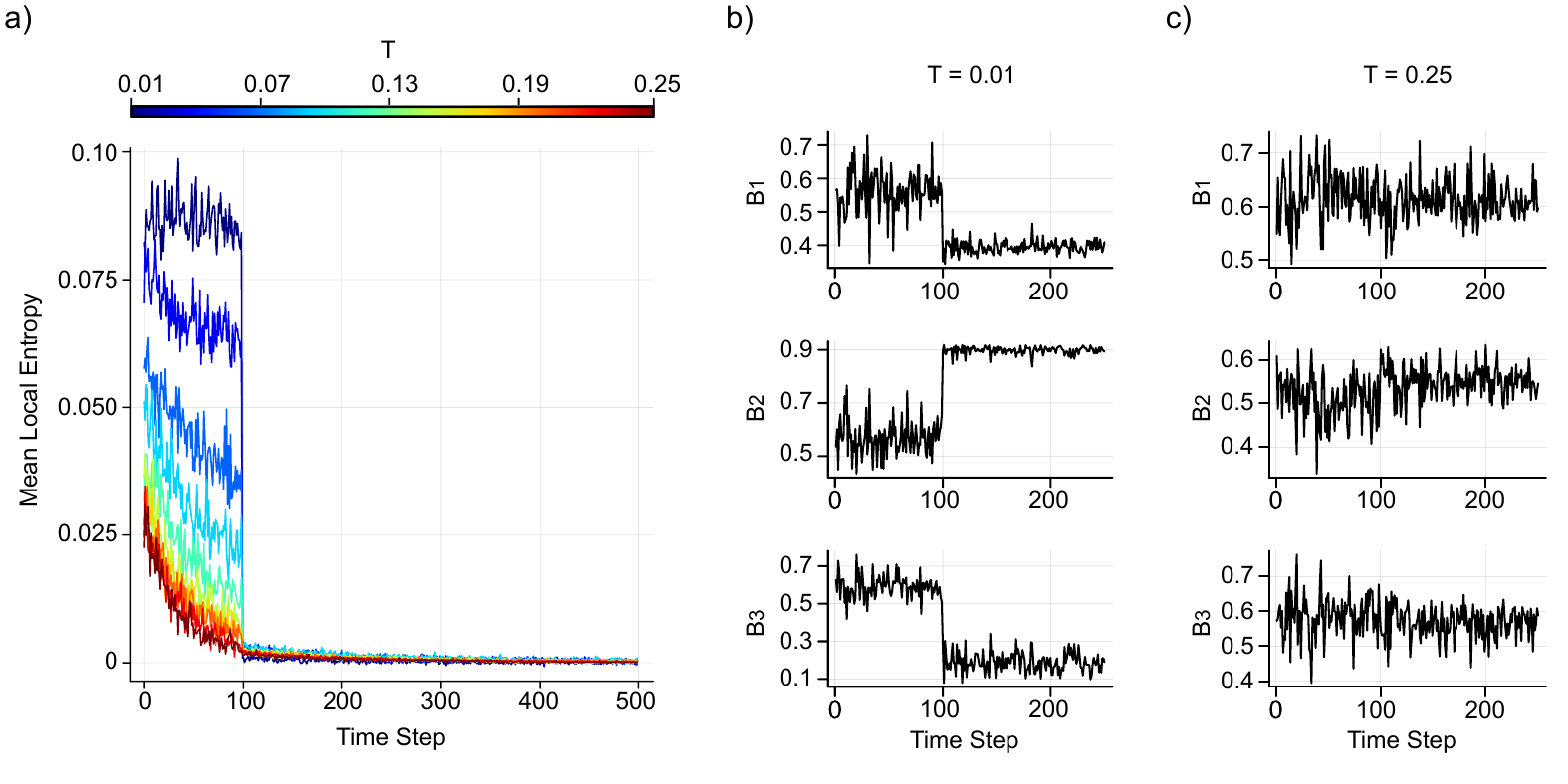}
    \caption{(a) Mean Local Entropy of the proposed model at different final temperatures, averaged over $500$ simulations, setting $M = 50$, $\lambda = 1.0$, and $G = 0.2$. The initial system temperature is equal to $T = 1000$ and is abruptly reduced to smaller values at the time step $t = 100$. The two plots on the right show the outcomes of each single strategy component, forming the block, for the lowest (b) and highest (c) value of temperature here considered, i.e. $T = 0.01$ and $T = 0.25$, respectively.}
    \label{fig:fig_temp}
\end{figure}

\begin{figure}
    \centering
    \includegraphics[width=180mm]{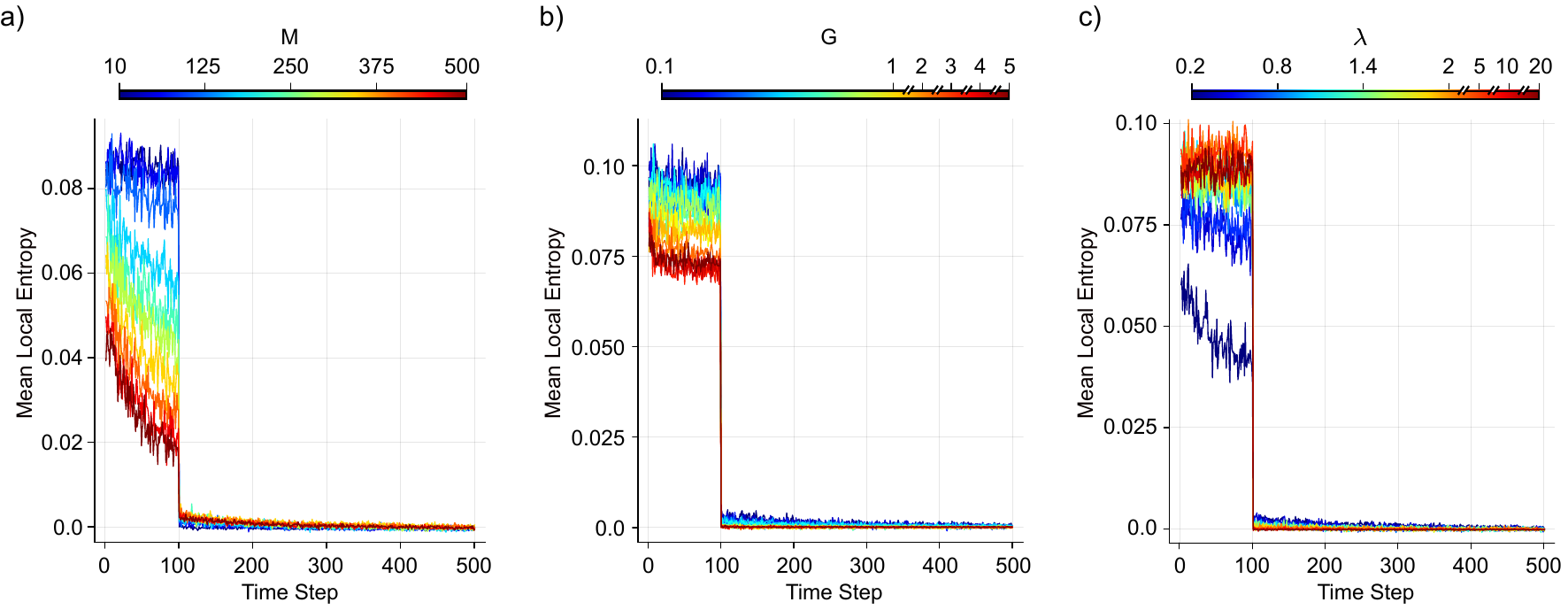}
    \caption{The entropy of the agent population on varying the number of strategies $M$, the growth factor $G$ and the non-linearity $\lambda$ of payoff distribution. We performed 500 independent simulations and computed the mean of the local entropy as a function of time: (a) Results related to the number of strategies setting $T = 0.01$, $\lambda = 1.0$, $G = 0.2$; (b) Results related to the linear growth factors setting $T = 0.01$, $\lambda=1.0$, $M = 50$; (c) Results related to different values of $\lambda$, setting $T = 0.01$, $M = 50$, $G = 0.2$. Block state fluctuations are regularised by the population growth, leading the local entropy to decrease over time for small values of $\lambda$ or high values of $M$. Such effect is less relevant for large values of $\lambda$, and the whole population converges towards a common behaviour.}
    \label{fig:figs}
\end{figure}

Each strategy of the proposed game is associated with a payoff that, in suitable conditions, can affect the strategy selection agents perform. Therefore, the generation and association of payoffs to strategies have a fundamental role in these dynamics.
For instance, a homogeneous payoff distribution entails that the strategy selection is equivalent to a random process. 
Similarly, payoff heterogeneity can trigger agents to choose the most convenient strategies.
Here, we generate payoffs by an exponential curve. This choice yields a heterogeneous payoff distribution and allows us to limit the number of strategies associated with high payoffs.
Figure~\ref{fig:payoff} shows the generation of $10$ strategies, each with $3$ components whose value ranges in $[0,1]$.
In the described scenario, a rational attitude promotes only the subset of convenient strategies.
That can be studied in detail by analysing the effects of the system temperature $T$, representing the agents' rationality, on the strategy selection process. 
We expect that at low temperatures, only convenient strategies get selected. While, at high temperatures, despite the payoff heterogeneity, all strategies get selected with the same probability, no matter their payoff.
Rationality is introduced in the agent population by an abrupt decrease in the system temperature.
More in detail, the initial system temperature is $T(0) = 1000$ then, at $t= 100$, it is abrouptly decreased to very low values, e.g. $T(100) = 0.1$.
In summary, tuning the value of $T$ allows us to increase/decrease the rational attitude among agents.
Results of the $M$-strategy game can be analysed through the entropy $S$ as defined in equation~\ref{eq:entropy_measure}. Accordingly, Figure~\ref{fig:fig_temp} shows the entropy $S$ in function of the temperature $T$. 
As reported in plot \textbf{c} of Figure~\ref{fig:fig_temp}, if $T(100)$ is equal or greater than $0.25$, which is a value relatively very small compared to $T(0)$, the block components $B_1$, $B_2$ and $B_3$ are still selected almost uniformly to form a strategy. 
On the other hand, for $T(100) \to 0.01$, the three blocks forming strategies reach a constant value, which results in a sharp decrease in the entropy $S$.
Next, we investigate the entropy on varying the non-linearity $\lambda$ of the distribution used for generating payoffs, the number of strategies $M$ in the model, and the growth factor $G$ governing the number of new agents added in the population at each step ---see Figure~\ref{fig:figs}. These analyses are performed by setting $T = 0.01$.
All considered model parameters affect the strategy profile. Starting with the number of strategies (plot \textbf{a} of Figure~\ref{fig:figs}), even when $M$ increases, the transition from irrational to rational behaviour still occurs, although the difference between $S(t<100)$ and $S(t\ge100)$ reduces. As the number of strategies available to agents increases, so does the pool of high-yield options. That gives a scenario similar to that obtained at high temperatures, with multiple high payoff strategies.
Then, looking at the growth factor $G$ (plot \textbf{b} of Figure~\ref{fig:figs}), increasing the population size over time reflects on a reduced difference between the initial and the final entropy for the chosen setting.
Unlike the other parameters, the growth factor does not influence the decision-making process of a single agent. On the contrary, it alters the degree of variability of the block state, which results from the collective behaviour of all users. An increase in growth factor leads to faster stabilisation of average user behaviour.
Eventually, on varying $\lambda$ (whose effect has been, in part, described above), we find that the higher this parameter, the higher the difference between the initial and the final entropy, making relevant the irrational to rational transition in the population.
\subsection*{Numerical Simulations vs Bitcoin Market}
\begin{figure}
    \centering
   \includegraphics[width=180mm]{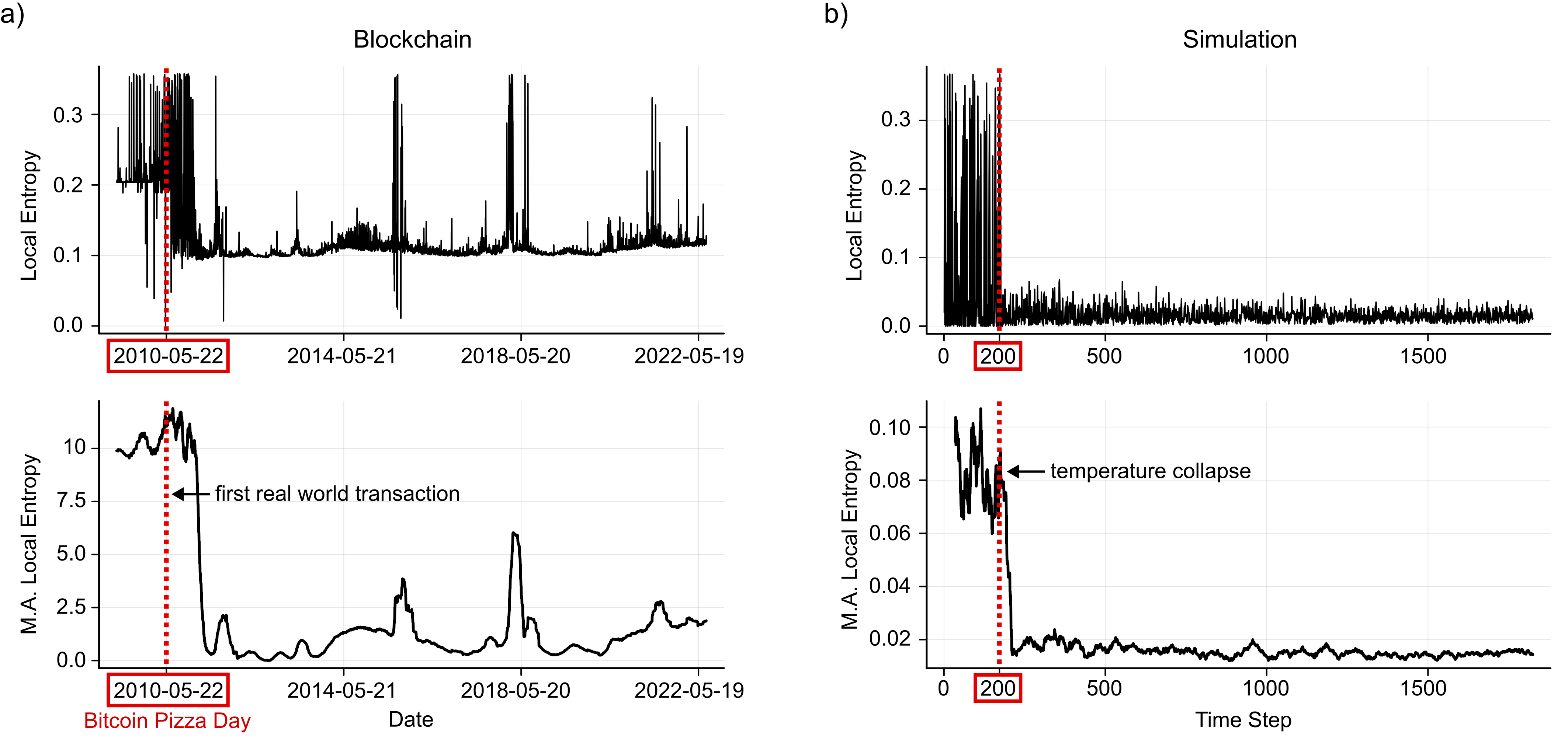}
    \caption{(a) Local entropy computed of the Blockchain. (b) Local entropy of the agent population. The latter is derived by setting $T = 0.01$, $\lambda = 1.0$, $M = 50$, and $G = 0.2$. The top diagrams (black lines) show the local entropy at each time step, while the bottom diagrams show the moving average (mean over previous values) of the related signal on time windows of $90$ days and $35$ steps, respectively. 
The drop in entropy occurring in the Blockchain, following a slight increase in the market value of BTC resembles the drop in entropy observed in the agent model resulting from an abrupt reduction of the system temperature $T$.}
   \label{fig:comparison_entropy}
\end{figure}

\begin{figure}
    \centering
   \includegraphics[width=176mm]{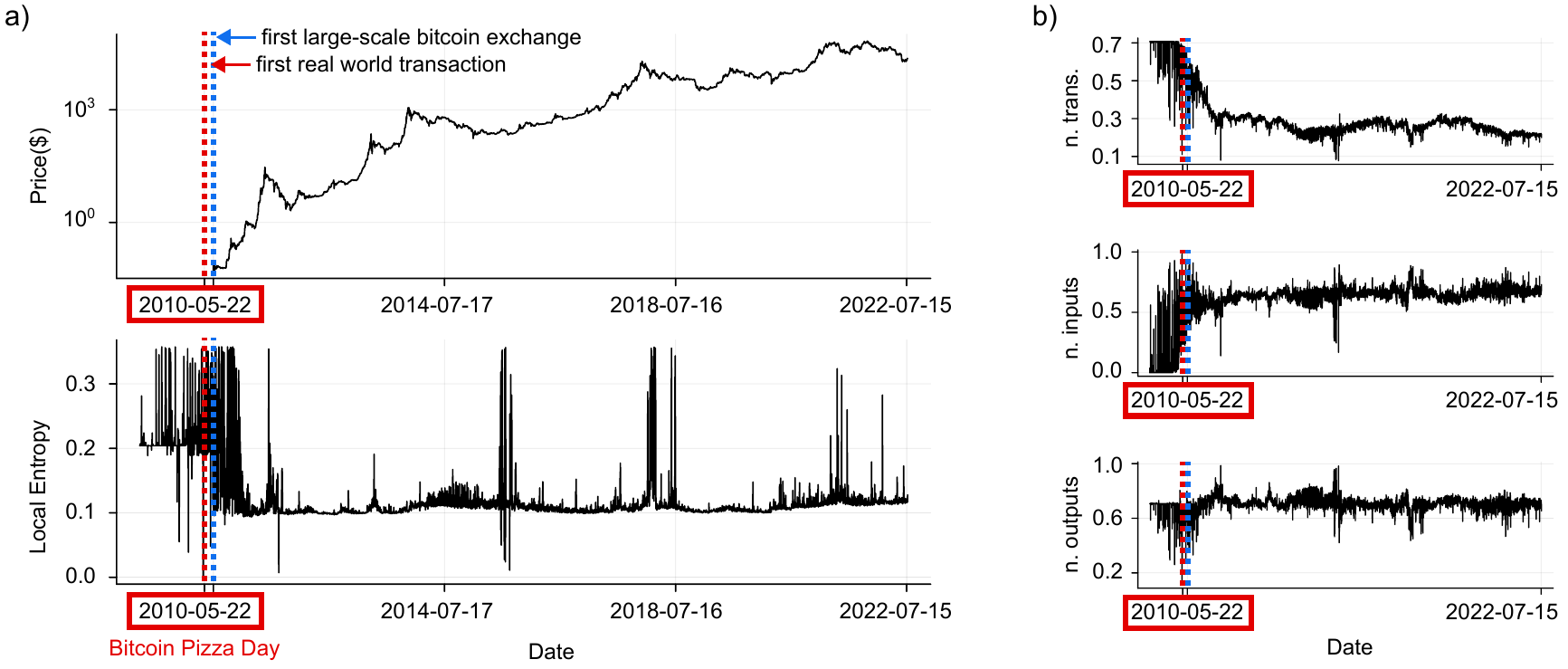}
    \caption{(a) On the top, the price history of BTC\slash USD in log-scale, and on the bottom, the local entropy in time. As the number of users increases and people become more aware of the value of crypto, entropy decreases in a relatively short time. A below-average value is a symptom of a rare event, e.g. the Flood Attack~\cite{flood01} that occurred in 2015 manifests as a relevant fluctuation. (b) Components of the normalized block state over time: number of transactions, number of inputs, and number of outputs. The variables, reflecting the user's behaviour, gradually stabilise.}
    \label{fig:price}
\end{figure}

The most suitable values of the model parameters, leading to a behavioural transition in the agent population, are used to perform a comparison with the Bitcoin users' behaviour.
As above-described, each block of the Blockchain gets mapped to a vector, i.e. by considering the number of inputs, number of outputs, and the number of Bitcoin transactions it contains. Such mapping allows us to compute the entropy in the Blockchain.
Results are illustrated in Figure~\ref{fig:comparison_entropy}. The latter shows that the two populations, i.e. the synthetic one (composed of agents) and the real one (composed of Bitcoin users), have very similar behaviour.
In light of that, we look at the $BTC\slash USD$ in Figure~\ref{fig:price} signal to assess whether it might have a role in the entropy transition detected on the Blockchain. The relationships between Blockchain dynamics and the crypto market still needs to be clarified, yet, some preliminary studies (e.g.~\cite{javarone01}) highlight interesting interactions between these two systems.
As performed for the single components of blocks, in the agent population, plot \textbf{b} of Figure~\ref{fig:price} shows the variation of the three parameters in the Blockchain. In particular, the normalized number of transactions and that of inputs converge towards a stable trend.
In summary, the temperature collapse in the agent-based model seems to be related to some critical events that occurred on the Blockchain in the period May-July 2010. In plot \textbf{a} of Figure~\ref{fig:price}, we highlight the famous Pizza Day~\cite {pizza_day}. In addition, the first official Bitcoin exchange, Mt. Gox, launched in the same period.
\section{Discussion and Conclusion}\label{sec:conclusion} 
Rationality is a fundamental ingredient in many decision processes, such as those required to manage resources and financial assets.
Usually, economic agents are considered rational, yet detecting and measuring rationality in individuals and communities is a non-trivial challenge.
To study such relevant behavioural attitude, we focused on cryptocurrencies (cryptos), i.e. an innovative application resulting from the Blockchain.
In the last years, this application led to the emergence of a growing system, namely the crypto market (e.g. see~\cite{pessa01,ding01,stosic01,stosic02}), whose dynamics are as complex as those governing stock markets, forex, and similar platforms.
As mentioned, Bitcoin is the first cryptocurrency. Its growth is particularly interesting for understanding how the whole crypto market rose, at least in its earliest times.
Here, we hypothesise that the dynamics underlying Bitcoin, and the few cryptos launched right after its circulation, contain information to detect hallmarks of rationality among crypto users/owners. 
Our hypothesis relies on the following observations. Firstly, Bitcoin experienced a rise from $0$ to more than $60$ thousand US dollars in a very few years.
Reasonably, in this period, the attitude of Bitcoin owners in managing their crypto wallets may have widely changed. The Pizza Day~\cite {pizza_day} fairly reflects a relaxed way to deal with Bitcoin, i.e. as a token attracting the curiosity of geeks and a few other individuals. 
On the other hand, as soon as BTC rose in the market to hundreds and even thousands of US dollars, wallet owners had to put much more care into making transactions.
Therefore, we assume that an initial financially riskless electronic item that soon became a high-value asset can constitute a valuable resource for investigating human rationality. Also, Bitcoin transactions are available online, so the data we need can be directly accessed and analysed.
In order to test our hypothesis and study the rationality of Bitcoin users, we implemented the following method. 
First, we start from a simple agent-based model that includes a numerical parameter devised for controlling the rationality of agents. The proposed model allows us to observe whether the agent behaviour has similarities with the Bitcoin users' behaviour. To this end, we adopt an entropy measure.
Results show a sharp transition of the entropy in both populations, i.e. the synthetic one and the real one composed of Bitcoin users.
The dynamics of Bitcoin users can get inferred by the data recorded in the Blockchain, mapping users' actions (in performing a transaction) to strategies of a generic game. 
The strategy selection process of this game has a temperature (noise) parameter. Accordingly, we can analyse the strategy distribution by varying this temperature, representing the degree of rationality.
Remarkably we observe a sharp entropy transition in the Blockchain, which coincides with a modest increase in the Bitcoin quotation. Interestingly, even a market value slightly higher than zero is enough to trigger individuals to become rational. Then, to observe a similar phenomenon in the agent population, the system temperature has to be suddenly decreased to very low values.
This finding leads us to further considerations. We find confirmation that human attitude in managing resources changes drastically even when the value of these resources remains limited to a few dollars (or even less).
Then, rationality seems to rise as a global attitude, i.e. involving a growing community of individuals that, in principle, do not interact.
Ownership is the common element among these users, and despite the large spectrum of human behaviours, the market value of an asset seems to stimulate all individuals to act alike. 
The idea individuals aim to avoid losing money is fairly ascertained. Thus, our investigation confirms such a common idea and shows how fast a community changes its attitude toward managing a resource whose value grows.
In light of these results and previous investigation~\cite{javarone01}, we suggest Blockchain data can get exploited for studying relevant human behaviours. 
Among them, rationality is particularly interesting, and the entropy pattern we identified suggests this attitude can rise quite suddenly among a whole community regardless of its size.
To conclude, we deem our results shed light on the dynamics of human rationality, yet, further investigations can be highly beneficial to corroborate the validity of our findings.
\section*{Acknowledgement}
Authors wish to thank Marco Corradino for the stimulating discussions. MAJ is supported by the PNRR NQST (Code: $PE23$). 

\end{document}